\documentclass[12pt,a4paper]{article}
\usepackage{amsmath}
\usepackage{latexsym}
\makeatletter
\def\rddots{\mathinner{\mkern1mu\raise\p@%
    \vbox{\kern7\p@\hbox{.}}\mkern2mu%
    \raise4\p@\hbox{.}\mkern2mu\raise7\p@\hbox{.}\mkern1mu}}
\makeatother
\newcommand{\ket}[1]{{\vert{#1}\rangle}}

\newcommand{\fukuso}{{\mathbf C}}

\newcommand{\tr}{{\rm tr}}

\newcommand{\zetta}{{\vert z\vert}}

\begin{document}

\title{\sl Riccati Diagonalization of Hermitian Matrices}
\author{
  Kazuyuki FUJII
  \thanks{E-mail address : fujii@yokohama-cu.ac.jp }\quad and\ 
  Hiroshi OIKE
  \thanks{E-mail address : oike@tea.ocn.ne.jp }\\
  ${}^{*}$Department of Mathematical Sciences\\
  Yokohama City University\\
  Yokohama, 236--0027\\
  Japan\\
  ${}^{\dagger}$Takado\ 85--5,\ Yamagata, 990--2464\\
  Japan\\
  }
\date{}
\maketitle
\begin{abstract}
  In this paper a geometric method based on Grassmann manifolds 
and matrix Riccati equations to make hermitian matrices diagonal is 
presented. We call it Riccati Diagonalization.
\end{abstract}

\section{Introduction}
In this paper we consider a finite dimensional quantum model, 
so its Hamiltonian is a (finite-dimensional) hermitian matrix.  
In order to solve the model we want to make the Hamiltonian diagonal.

Although we have a standard method for the purpose, to perform it 
{\bf explicitly} is very hard (maybe, almost impossible). 
Let us make a brief introduction, see \cite{Sa}, \cite{St}.

Let $H$ be a hermitian matrix, namely
\begin{equation}
H\in \{H\in M(n;\fukuso)\ |\ H^{\dagger}=H\}.
\end{equation}
When we want to make $H$ diagonal 
Elementary Linear Algebra shows 
the following diagonalization procedure 
$[A]\Longrightarrow [B]\Longrightarrow [C]$ :

\noindent
[A]\ \ First we calculate eigenvalues of $H$,
\begin{equation}
0
=|\lambda E-H|
=\lambda^{n}-\tr{H}\lambda^{n-1}+\cdots +(-1)^{n}\det{H}.
\end{equation}
There are $n$ real solutions although it is almost impossible to 
look for exact ones, so let those be 
$\{\lambda_{1}, \lambda_{2}, \cdots, \lambda_{n}\}$.

\noindent
[B]\ \ Next we find eigenvectors $\{\ket{\lambda}\}$ corresponding 
to eigenvalues
\begin{equation}
H\ket{\lambda_{j}}=\lambda_{j}\ket{\lambda_{j}}
\quad \mbox{and} \quad 
\langle{\lambda_{i}}|{\lambda_{j}}\rangle=\delta_{ij}
\end{equation}
for all $1\leq i,\ j\leq n$. It is also almost impossible to 
carry out.

\noindent
[C]\ \ Last by setting
\begin{equation}
U=(\ket{\lambda_{1}},\ket{\lambda_{2}},\cdots,\ket{\lambda_{n}})
\end{equation}
we finally obtain
\begin{equation}
H=UD_{H}U^{\dagger}
\end{equation}
where $D_{H}=
\mbox{diag}(\lambda_{1},\lambda_{2},\cdots,\lambda_{n})$ is the 
diagonal matrix.

The procedure is standard, while to carry out it completely is 
another problem\footnote{in Japan it is called a pie in the sky}.
Even if $n=3$ it is very hard. In fact, for the hermitian matrix
\[
H=
\left(
\begin{array}{ccc}
h_{1} & \bar{\alpha} & \bar{\beta} \\
\alpha  & h_{2} & \bar{\gamma} \\
\beta  & \gamma & h_{3}
\end{array}
\right)\in H(3;\fukuso)
\]
carry out the diagonalization. As far as we know it has not 
been given in any textbook on linear algebra.

Note that the characteristic equation 
$f(\lambda)=|\lambda E-H|$ is given by
\begin{eqnarray*}
f(\lambda)
&=&
\lambda^{3}-(h_{1}+h_{2}+h_{2})\lambda^{2}+
(h_{1}h_{2}+h_{1}h_{3}+h_{2}h_{3}-|\alpha|^{2}-|\beta|^{2}-|\gamma|^{2})
\lambda+ \\
&&|\gamma|^{2}h_{1}+|\beta|^{2}h_{2}+|\alpha|^{2}h_{3}-h_{1}h_{2}h_{3}
-\alpha\bar{\beta}\gamma-\bar{\alpha}\beta\bar{\gamma}.
\end{eqnarray*}
To look for exact solutions by use of Cardano formula is not easy 
(for example, try it by use of MATHEMATICA or MAPLE).

Therefore we present another diagonalization method based on 
{\bf Grassmann manifolds} and {\bf matrix Riccati equations}.

\section{Riccati Diagonalization}
In this section we show a new diagonalization method. Before it 
let us explain the idea with simple example for beginners. 
The target is
\begin{equation}
H=
\left(
\begin{array}{cc}
h_{1}     & \bar{\alpha} \\
\alpha  & h_{2}        
\end{array}
\right)\in H(2;\fukuso).
\end{equation}

To diagonalize $H$ above we consider a matrix 
\begin{equation}
U\equiv U(z)=\frac{1}{\sqrt{1+\zetta^{2}}}
\left(
\begin{array}{cc}
1  & -\bar{z} \\
z  & 1        
\end{array}
\right)\quad \mbox{for}\quad z\in \fukuso.
\end{equation}
It is easy to check $U^{\dagger}U=UU^{\dagger}=1_{2}$ and 
$|U|=1$, so $U$ is special unitary ($U\in SU(2)$). Namely, 
$U$ is a map
\[
U\ :\ \fukuso\ (\subset S^{2})\ \longrightarrow\ SU(2).
\]
This map is well--known in Mathematics or Mathematical Physics.

The calculation $U^{\dagger}HU$ gives
\begin{eqnarray}
U^{\dagger}HU
&=&
\frac{1}{1+\zetta^{2}}
\left(
\begin{array}{cc}
1  & \bar{z} \\
-z & 1        
\end{array}
\right)
\left(
\begin{array}{cc}
h_{1}     & \bar{\alpha} \\
\alpha  & h_{2}        
\end{array}
\right)
\left(
\begin{array}{cc}
1  & -\bar{z} \\
z  &   1        
\end{array}
\right) \nonumber \\
&=&
\frac{1}{1+\zetta^{2}}
\left(
\begin{array}{cc}
h_{1}+\bar{\alpha}z+\alpha\bar{z}+h_{2}\zetta^{2}  &
\bar{\alpha}-(h_{1}-h_{2})\bar{z}-\alpha\bar{z}^{2}    \\
\alpha -(h_{1}-h_{2})z-\bar{\alpha}z^{2} &
h_{2}-\bar{\alpha}z-\alpha\bar{z}+h_{1}\zetta^{2}
\end{array}
\right),
\end{eqnarray}
so if we assume the equation
\begin{equation}
\label{eq:riccati}
\alpha-(h_{1}-h_{2})z-\bar{\alpha}z^{2} =0
\Longleftrightarrow 
\bar{\alpha}z^{2}+(h_{1}-h_{2})z-\alpha =0
\end{equation}
we have the diagonal matrix
\begin{equation}
U^{\dagger}HU
=
\left(
\begin{array}{cc}
\frac{h_{1}+\bar{\alpha}z+\alpha\bar{z}+h_{2}\zetta^{2}}
{1+\zetta^{2}} &    \\
    &
\frac{h_{2}-\bar{\alpha}z-\alpha\bar{z}+h_{1}\zetta^{2}}
{1+\zetta^{2}}
\end{array}
\right).
\end{equation}
From this two eigenvalues are obtained by
\begin{equation}
\lambda_{1}=\frac{h_{1}+\bar{\alpha}z+\alpha\bar{z}+h_{2}\zetta^{2}}
{1+\zetta^{2}},\ \ 
\lambda_{2}=\frac{h_{2}-\bar{\alpha}z-\alpha\bar{z}+h_{1}\zetta^{2}}
{1+\zetta^{2}}
\end{equation}
under the equation (\ref{eq:riccati}), whose solutions are easily given by
\begin{equation}
z=\frac{-(h_{1}-h_{2})\pm\sqrt{(h_{1}-h_{2})^{2}+4|\alpha|^{2}}}{2\bar{\alpha}}.
\end{equation}

\vspace{3mm}
A comment is in order. The equation (\ref{eq:riccati}) is a special version 
of  (generalized) Riccati equations.

As shown in the example our diagonalization method is different from 
usual one (a kind of reverse procedure). Let us state our procedure.

\vspace{5mm}\noindent
{\bf Riccati Diagonalization}

\noindent
[A]\ \ For $H\in H(n;\fukuso)$ we prepare a unitary matrix 
$U=U(Z)\in U(n)$ where $Z$ is a parameter matrix and calculate
\[ 
U^{\dagger}HU\equiv W=(w_{ij}).
\]

\noindent
[B]\ \ We set 
\[
w_{ij}=0\quad \mbox{for}\quad 1\leq j < i \leq n
\]
and solve these simultaneous equations (a system of Riccati equations) 
to determine $Z=(z_{kl})$.

\noindent
[C]\ \ We finally obtain the diagonal matrix
\[
W=\mbox{diag}(w_{11}, w_{22},\cdots, w_{nn})
\]
where each component is an eigenvalue of $H$ under [B].

\section{General Case}

In this section we consider a generalization of the example 
in the preceding section.  See \cite{Oi}, \cite{Fu1} as a 
general introduction to Grassmann manifolds and also 
\cite{Ga} and its references as an application.

Namely, we treat a hermitian matrix
\begin{equation}
H=
\left(
\begin{array}{cc}
H_{+} & V^{\dagger}  \\
V & H_{-}
\end{array}
\right)\in H(n;\fukuso)
\end{equation}
where
\[
H_{+}\in H(k;\fukuso),\quad
H_{-}\in H(n-k;\fukuso),\quad
V\in M(n-k,k;\fukuso)
\]
for $1\leq k\leq n-1$. In order to make $H$ a direct sum form 
we prepare a unitary matrix
\begin{eqnarray}
U&=&U(Z)=
\left(
\begin{array}{cc}
1_{k} & -Z^{\dagger}  \\
Z & 1_{n-k}
\end{array}
\right)
\left(
\begin{array}{cc}
(1_{k}+Z^{\dagger}Z)^{-1/2} &                                          \\
                                     & (1_{n-k}+ZZ^{\dagger})^{-1/2} 
\end{array}
\right) \\
&\equiv&U_{M}U_{D}  \nonumber
\end{eqnarray}
where $Z\in M(n-k,k;\fukuso)$ is a parameter matrix. 
$U$ is a map
\[
U\ :\ M(n-k,k;\fukuso)\longrightarrow\ SU(n)
\]
and $Z$ is a local coordinate of the Grassmann manifold 
$G_{k}(\fukuso^{n})$ defined by
\begin{eqnarray*}
G_{k}(\fukuso^{n})
&=&
\{P\in M(n;\fukuso)\ |\ P^{2}=P,\ P^{\dagger}=P,\ \mbox{tr}P=k\} \\
&=&
\{UP_{0}U^{\dagger}\ |\ U\in U(n)\} \\
&\cong & U(n)/U(k)\times U(n-k)
\end{eqnarray*}
with $P_{0}$ given by
\[
P_{0}=
\left(
\begin{array}{cc}
1_{k} &            \\
       & 0_{n-k}
\end{array}
\right).
\]
Note that $\mbox{dim}_{\fukuso}G_{k}(\fukuso^{n})=
k(n-k)=\mbox{dim}_{\fukuso}M(n-k,k;\fukuso)$. 
The local parametrization of $G_{k}(\fukuso^{n})$ is 
more explicitly given by
\begin{equation}
P(Z)=U(Z)P_{0}U(Z)^{\dagger}
=
\left(
\begin{array}{cc}
1_{k} & -Z^{\dagger}  \\
Z & 1_{n-k}
\end{array}
\right)
\left(
\begin{array}{cc}
1_{k} &            \\
       & 0_{n-k}
\end{array}
\right)
\left(
\begin{array}{cc}
1_{k} & -Z^{\dagger}  \\
Z & 1_{n-k}
\end{array}
\right)^{-1}
\end{equation}
where we have used the relation
\[
\left(
\begin{array}{cc}
1_{k} & -Z^{\dagger}  \\
Z & 1_{n-k}
\end{array}
\right)^{-1}
=
\left(
\begin{array}{cc}
(1_{k}+Z^{\dagger}Z)^{-1} &                                      \\
                                 & (1_{n-k}+ZZ^{\dagger})^{-1} 
\end{array}
\right)
\left(
\begin{array}{cc}
1_{k} & -Z^{\dagger}  \\
Z & 1_{n-k}
\end{array}
\right)^{\dagger}.
\]

Let us calculate $U_{M}^{\dagger}HU_{M}$ :
\begin{eqnarray}
U_{M}^{\dagger}HU_{M}
&=&
\left(
\begin{array}{cc}
1_{k} & Z^{\dagger}  \\
-Z & 1_{n-k}
\end{array}
\right)
\left(
\begin{array}{cc}
H_{+} & V^{\dagger}  \\
V & H_{-}
\end{array}
\right)
\left(
\begin{array}{cc}
1_{k} & -Z^{\dagger}  \\
Z & 1_{n-k}
\end{array}
\right)  \nonumber \\
&=&
\left(
\begin{array}{cc}
H_{+}+Z^{\dagger}V+V^{\dagger}Z+Z^{\dagger}H_{-}Z &
V^{\dagger}-H_{+}Z^{\dagger}+Z^{\dagger}H_{-}-Z^{\dagger}VZ^{\dagger} \\
V-ZH_{+}+H_{-}Z-ZV^{\dagger}Z & 
H_{-}-ZV^{\dagger}-VZ^{\dagger}+ZH_{+}Z^{\dagger}
\end{array}
\right).
\end{eqnarray}
From this we set
\begin{equation}
\label{eq:matrix Riccati equation}
V-ZH_{+}+H_{-}Z-ZV^{\dagger}Z=0\ \Longleftrightarrow\ 
ZV^{\dagger}Z+ZH_{+}-H_{-}Z-V=0.
\end{equation}
This is just the matrix Riccati equation. Under the condition we obtain 
the block form 
\begin{equation}
U^{\dagger}HU
=
\left(
\begin{array}{cc}
(1_{k}+Z^{\dagger}Z)^{-1/2}\widetilde{H}_{+}(1_{k}+Z^{\dagger}Z)^{-1/2} &         \\
  & (1_{n-k}+ZZ^{\dagger})^{-1/2}\widetilde{H}_{-}(1_{n-k}+ZZ^{\dagger})^{-1/2} 
\end{array}
\right)
\end{equation}
where
\begin{equation}
\label{eq:reduced Hamiltonian}
\widetilde{H}_{+}=H_{+}+Z^{\dagger}V+V^{\dagger}Z+Z^{\dagger}H_{-}Z,\quad 
\widetilde{H}_{-}=H_{-}-ZV^{\dagger}-VZ^{\dagger}+ZH_{+}Z^{\dagger}.
\end{equation}

How to solve the Riccati equation (\ref{eq:matrix Riccati equation}) 
is not known  as far as we know. In fact, it is very hard, so we must 
satisfy by finding some approximate solution at the present time. 

\vspace{3mm} \noindent
{\bf Approximation I}

First, by rejecting the quadratic term we have
\begin{equation}
\label{eq:linear Riccati equation}
ZH_{+}-H_{-}Z=V.
\end{equation}
This solution is well--known to become
\begin{equation}
\label{eq:linear solution}
Z=\int_{0}^{\infty}e^{tH_{-}}Ve^{-tH_{+}}dt
\end{equation}
under some condition on $H_{-}$ and $H_{+}$. See for example 
\cite{Ari}. In fact,
\begin{eqnarray*}
ZH_{+}-H_{-}Z
&=&
\int_{0}^{\infty}
\{e^{tH_{-}}Ve^{-tH_{+}}H_{+}-H_{-}e^{tH_{-}}Ve^{-tH_{+}}\}dt \\
&=&
-\int_{0}^{\infty}\frac{d}{dt}(e^{tH_{-}}Ve^{-tH_{+}})dt \\
&=&
-[e^{tH_{-}}Ve^{-tH_{+}}]_{0}^{\infty} \\
&=&
V
\end{eqnarray*}
under the condition
\begin{equation}
\lim_{t\rightarrow \infty}e^{tH_{-}}Ve^{-tH_{+}}=0.
\end{equation}

\vspace{3mm} \noindent
{\bf Approximation II}

Next, let us consider another approximation. We assume that 
$n=2m,\ k=m$ and $V$ is invertible ($V\in GL(m;\fukuso)$). 
Then, by remembering
\[
ax^{2}+2bx+c=0
\ \Longrightarrow \ 
a(x+\frac{b}{a})^{2}=-c+\frac{b^{2}}{a}
\]
we have
\begin{equation}
\{Z-H_{-}(V^{\dagger})^{-1}\}V^{\dagger}\{Z+(V^{\dagger})^{-1}H_{+}\}
=
V-H_{-}(V^{\dagger})^{-1}H_{+}
\end{equation}
from (\ref{eq:matrix Riccati equation}). Here if we can choose $Z$ as 
\[
Z+(V^{\dagger})^{-1}H_{+}\in GL(m;\fukuso)
\]
then we have a recursive relation
\begin{equation}
Z=H_{-}(V^{\dagger})^{-1}+
\{V-H_{-}(V^{\dagger})^{-1}H_{+}\}
\frac{1}{Z+(V^{\dagger})^{-1}H_{+}}(V^{\dagger})^{-1}.
\end{equation}
Now by inserting an approximate solution (\ref{eq:linear solution}) into 
the equation above we obtain the approximate solution
\begin{equation}
Z\approx H_{-}(V^{\dagger})^{-1}+
\{V-H_{-}(V^{\dagger})^{-1}H_{+}\}
\frac{1}{
\int_{0}^{\infty}e^{tH_{-}}Ve^{-tH_{+}}dt+
(V^{\dagger})^{-1}H_{+}}
(V^{\dagger})^{-1}.
\end{equation}
if 
\[
\int_{0}^{\infty}e^{tH_{-}}Ve^{-tH_{+}}dt+(V^{\dagger})^{-1}H_{+}
\in GL(m;\fukuso)
\]
or 
\[
H_{+}+\int_{0}^{\infty}V^{\dagger}e^{tH_{-}}Ve^{-tH_{+}}dt
\in GL(m;\fukuso).
\]

A comment is in order.\ \ We don't know at the present time 
whether our approximate solution is convenient enough or not.

\section{Reduction of Riccati Diagonalization}
In this section we give an explicit procedure of Riccati diagonalization. 
General Hamiltonian is
\begin{equation}
H=
\left(
\begin{array}{ccccccc}
h_{1} & \bar{v}_{21} & \bar{v}_{31} & \cdot & \cdot & \bar{v}_{n-1,1} & \bar{v}_{n1}  \\
v_{21} & h_{2} & \bar{v}_{32} & \cdot & \cdot & \bar{v}_{n-1,2} & \bar{v}_{n2}          \\
v_{31} & v_{32} & h_{3} & \cdot & \cdot & \bar{v}_{n-1,3} & \bar{v}_{n3}                  \\
\cdot & \cdot & \cdot & \cdot &    & \cdot  & \cdot                                       \\
\cdot & \cdot & \cdot &  & \cdot & \cdot & \cdot                                          \\
v_{n-1,1} & v_{n-1,2} & v_{n-1,3} & \cdot & \cdot & h_{n-1} & \bar{v}_{n,n-1}           \\
v_{n1} & v_{n2} & v_{n3} & \cdot & \cdot & v_{n,n-1} & h_{n}
\end{array}
\right)\ \in\ H(n;\fukuso)
\end{equation}
and we write as
\begin{equation}
H=
\left(
\begin{array}{cc}
H_{+} & V^{\dagger} \\
V     & h_{n}
\end{array}
\right),\quad
V=(v_{n1},v_{n2}, \cdots, v_{n,n-1})
\end{equation}
for simplicity. We prepare a unitary matrix
\begin{equation}
U=
\left(
\begin{array}{cc}
1_{n-1} & -Z^{\dagger} \\
Z        & 1
\end{array}
\right)
\left(
\begin{array}{cc}
(1_{n-1}+Z^{\dagger}Z)^{-1/2} &    \\
        & (1+ZZ^{\dagger})^{-1/2}
\end{array}
\right)
\end{equation}
where $Z=(z_{1},z_{2},\cdots,z_{n-1})$. Then the Riccati equation 
is
\begin{eqnarray}
ZV^{\dagger}Z+ZH_{+}-h_{n}Z-V=0
&\Longleftrightarrow&
\left(\sum_{j=1}^{n-1}\bar{v}_{nj}z_{j}\right)z_{k}+
\sum_{j=1}^{n-1}(H_{+})_{jk}z_{j}-h_{n}z_{k}-v_{nk}=0   \nonumber \\
&&\quad \mbox{for}\ \ 1\leq k\leq n-1. 
\end{eqnarray}

Note that to solve the equation(s) above explicitly is very hard, so 
in general we must satisfy by constructing some approximate solution.

If we can solve the equation(s) then
\begin{eqnarray}
U^{\dagger}HU
&=&
\left(
\begin{array}{cc}
(1_{n-1}+Z^{\dagger}Z)^{-1/2}\widetilde{H}_{+}(1_{n-1}+Z^{\dagger}Z)^{-1/2} &   \\
  &  \frac{\widetilde{h}_{n}}{1+\sum_{j=1}^{n-1}|z_{j}|^{2}}
\end{array}
\right) \\
\tilde{h}_{n}
&=&h_{n}-\sum_{j=1}^{n-1}(z_{j}\bar{v}_{nj}+c.c.)+
\sum_{j=1}^{n-1}\sum_{k=1}^{n-1}z_{j}(H_{+})_{jk}\bar{z}_{k}  \nonumber
\end{eqnarray}
and the procedure is reduced to the calculation of
\[
(1_{n-1}+Z^{\dagger}Z)^{-1/2}\widetilde{H}_{+}
(1_{n-1}+Z^{\dagger}Z)^{-1/2}
\]
, so we must calculate the term $(1_{n-1}+Z^{\dagger}Z)^{-1/2}$ exactly. 

We write
\begin{eqnarray*}
&&Z
=(z_{1},z_{2},\cdots,z_{n-1})
=z_{1}(1,w_{2},\cdots,w_{n-1})
\equiv z_{1}(1,W),
\quad
w_{j}=z_{j}/z_{1}  \\
&&ZZ^{\dagger}=|z_{1}|^{2}(1+WW^{\dagger})=\sum_{j=1}^{n-1}|z_{j}|^{2}
\end{eqnarray*}
for simplicity. Then
\[
1_{n-1}+Z^{\dagger}Z
=
1_{n-1}+|z_{1}|^{2}
\left(
\begin{array}{cc}
1               &  W                  \\
W^{\dagger} &  W^{\dagger}W
\end{array}
\right)
\]
and a unitary matrix given by
\begin{equation}
\widetilde{U}
=
\left(
\begin{array}{cc}
1               & -W       \\
W^{\dagger} & 1_{n-2}
\end{array}
\right)
\left(
\begin{array}{cc}
(1+WW^{\dagger})^{-1/2} &                                           \\
                                  & (1_{n-2}+W^{\dagger}W)^{-1/2} 
\end{array}
\right)
\end{equation}
gives
\[
\widetilde{U}
\left(
\begin{array}{cc}
1 &            \\
   & 0_{n-2}
\end{array}
\right)
\widetilde{U}^{\dagger}
=
\frac{1}{1+WW^{\dagger}}
\left(
\begin{array}{cc}
1               & W                 \\
W^{\dagger} & W^{\dagger}W
\end{array}
\right).
\]
Therefore
\begin{eqnarray*}
1_{n-1}+Z^{\dagger}Z
&=&
1_{n-1}+|z_{1}|^{2}(1+WW^{\dagger})
\widetilde{U}
\left(
\begin{array}{cc}
1 &            \\
   & 0_{n-2}
\end{array}
\right)
\widetilde{U}^{\dagger}  \\
&=&
\widetilde{U}
\left\{
1_{n-1}+\sum_{j=1}^{n-1}|z_{j}|^{2}
\left(
\begin{array}{cc}
1 &            \\
   & 0_{n-2}
\end{array}
\right)
\right\}
\widetilde{U}^{\dagger} \\
&=&
\widetilde{U}
\left(
\begin{array}{cc}
1+\sum_{j=1}^{n-1}|z_{j}|^{2} &            \\
                                     & 1_{n-2}
\end{array}
\right)
\widetilde{U}^{\dagger} \\
\end{eqnarray*}
and we have
\begin{equation}
\left(1_{n-1}+Z^{\dagger}Z\right)^{-1/2}
=
\widetilde{U}
\left(
\begin{array}{cc}
\frac{1}{\sqrt{1+\sum_{j=1}^{n-1}|z_{j}|^{2}}} &            \\
                                                         & 1_{n-2}
\end{array}
\right)
\widetilde{U}^{\dagger}.
\end{equation}
As a result the reduced Hamiltonian is
\begin{eqnarray}
&&(1_{n-1}+Z^{\dagger}Z)^{-1/2}\widetilde{H}_{+}(1_{n-1}+Z^{\dagger}Z)^{-1/2} 
\nonumber \\
&=&
\widetilde{U}
\left(
\begin{array}{cc}
\frac{1}{\sqrt{1+\sum_{j=1}^{n-1}|z_{j}|^{2}}} &            \\
                                                         & 1_{n-2}
\end{array}
\right)
\widetilde{U}^{\dagger}
\widetilde{H}_{+}
\widetilde{U}
\left(
\begin{array}{cc}
\frac{1}{\sqrt{1+\sum_{j=1}^{n-1}|z_{j}|^{2}}} &            \\
                                                         & 1_{n-2}
\end{array}
\right)
\widetilde{U}^{\dagger}
\end{eqnarray}
and we obtain
\begin{eqnarray}
&&
U^{\dagger}HU  \nonumber \\
&=&
\left(
\begin{array}{cc}
\widetilde{U}
\left(
\begin{array}{cc}
\frac{1}{\sqrt{1+\sum_{j=1}^{n-1}|z_{j}|^{2}}} &            \\
                                                         & 1_{n-2}
\end{array}
\right)
\widetilde{U}^{\dagger}
\widetilde{H}_{+}
\widetilde{U}
\left(
\begin{array}{cc}
\frac{1}{\sqrt{1+\sum_{j=1}^{n-1}|z_{j}|^{2}}} &            \\
                                                         & 1_{n-2}
\end{array}
\right)
\widetilde{U}^{\dagger}  &                                                                     \\
                                & \frac{\widetilde{h}_{n}}{1+\sum_{j=1}^{n-1}|z_{j}|^{2}}
\end{array}
\right).    \nonumber \\
&&{}
\end{eqnarray}
We have only to continue the reduction process one after another.

A comment is in order.\ \ Note that it is not easy to calculate 
$(1_{n-k}+Z^{\dagger}Z)^{-1/2}$ for $Z\in M(k,n-k;\fukuso)$ and 
$2\leq k\leq n-2$.

\vspace{5mm} \noindent
{\bf Note}\ \ There is no need to calculate $(1_{n-2}+W^{\dagger}W)^{-1/2}$ 
in $\widetilde{U}$ because 
\begin{eqnarray*}
&&\widetilde{U}
\left(
\begin{array}{cc}
\frac{1}{\sqrt{1+\sum_{j=1}^{n-1}|z_{j}|^{2}}} &            \\
                                                         & 1_{n-2}
\end{array}
\right)
\widetilde{U}^{\dagger}  \\
&&
=
\left(
\begin{array}{cc}
1               & -W       \\
W^{\dagger} & 1_{n-2}
\end{array}
\right)
\left(
\begin{array}{cc}
\frac{1}{(1+WW^{\dagger})\sqrt{1+\sum_{j=1}^{n-1}|z_{j}|^{2}}} &  \\
                                       & (1_{n-2}+W^{\dagger}W)^{-1}
\end{array}
\right)
\left(
\begin{array}{cc}
1                 & W       \\
-W^{\dagger} & 1_{n-2}
\end{array}
\right)         \\
&&
=
\left(
\begin{array}{cc}
1               & -W       \\
W^{\dagger} & 1_{n-2}
\end{array}
\right)
\left(
\begin{array}{cc}
\frac{1}{(1+WW^{\dagger})\sqrt{1+\sum_{j=1}^{n-1}|z_{j}|^{2}}} &         \\
                        & 1_{n-2}-\frac{1}{1+WW^{\dagger}}W^{\dagger}W
\end{array}
\right)
\left(
\begin{array}{cc}
1                 & W       \\
-W^{\dagger} & 1_{n-2}
\end{array}
\right)
\end{eqnarray*}
from the definition of $\widetilde{U}$. This is important.

Last, let us make a comment. If $n=3$, namely $Z=(z_{1},z_{2})$ 
then we have a direct method (without using $W$). For
\[
1_{2}+Z^{\dagger}Z
=
\left(
\begin{array}{cc}
1+|z_{1}|^{2} & \bar{z}_{1}z_{2}  \\
z_{1}\bar{z}_{2} & 1+|z_{2}|^{2}
\end{array}
\right)
\]
we set
\[
U=
\frac{1}{\sqrt{|z_{1}|^{2}+|z_{2}|^{2}}}
\left(
\begin{array}{cc}
\bar{z}_{1}  & -z_{2}  \\
\bar{z}_{2}  & z_{1}
\end{array}
\right)\ \in\ SU(2).
\]
Then we have
\[
1_{2}+Z^{\dagger}Z
=
U
\left(
\begin{array}{cc}
1+|z_{1}|^{2}+|z_{2}|^{2} &   \\
                             & 1 
\end{array}
\right)
U^{\dagger}
\]
and
\begin{eqnarray}
(1_{2}+Z^{\dagger}Z)^{-1/2}
&=&
U
\left(
\begin{array}{cc}
\frac{1}{\sqrt{1+|z_{1}|^{2}+|z_{2}|^{2}}} &   \\
                                                  & 1 
\end{array}
\right)
U^{\dagger}  \nonumber \\
&=&
\frac{1}{|z_{1}|^{2}+|z_{2}|^{2}} 
\left(
\begin{array}{cc}
\bar{z}_{1}  & -z_{2}  \\
\bar{z}_{2}  & z_{1}
\end{array}
\right)
\left(
\begin{array}{cc}
\frac{1}{\sqrt{1+|z_{1}|^{2}+|z_{2}|^{2}}} &   \\
                                                  & 1 
\end{array}
\right)
\left(
\begin{array}{cc}
z_{1}           & z_{2}          \\
-\bar{z}_{2}  & \bar{z}_{1}
\end{array}
\right)  \nonumber \\
&=&
\frac{1}{|z_{1}|^{2}+|z_{2}|^{2}} 
\left(
\begin{array}{cc}
\frac{|z_{1}|^{2}}{\sqrt{1+|z_{1}|^{2}+|z_{2}|^{2}}}+|z_{2}|^{2}
&
\frac{\bar{z}_{1}z_{2}}{\sqrt{1+|z_{1}|^{2}+|z_{2}|^{2}}}-\bar{z}_{1}z_{2} \\
\frac{z_{1}\bar{z}_{2}}{\sqrt{1+|z_{1}|^{2}+|z_{2}|^{2}}}-z_{1}\bar{z}_{2}
&
\frac{|z_{2}|^{2}}{\sqrt{1+|z_{1}|^{2}+|z_{2}|^{2}}}+|z_{1}|^{2}
\end{array}
\right).
\end{eqnarray}
This form looks smart and will be used in the next section.

\section{Example}
In this section we apply our method to the following 
important example
\begin{equation}
H=
\left(
\begin{array}{ccc}
h_{1} & \bar{\alpha} & \bar{\beta} \\
\alpha  & h_{2} & \bar{\gamma} \\
\beta  & \gamma & h_{3}
\end{array}
\right).
\end{equation}
Since
\[
H_{+}=
\left(
\begin{array}{cc}
h_{1} & \bar{\alpha} \\
\alpha  & h_{2}
\end{array}
\right),
\quad
V=(\beta,\gamma),
\quad
Z=(z_{1},z_{2})
\]
the Riccati equation is
\[
\left\{
\begin{array}{ll}
\bar{\beta}z_{1}^{2}+\bar{\gamma}z_{1}z_{2}+(h_{1}z_{1}+\alpha z_{2})
-h_{3}z_{1}-\beta=0,  \\
\bar{\gamma}z_{2}^{2}+\bar{\beta}z_{1}z_{2}+(\bar{\alpha}z_{1}+h_{2}z_{2})
-h_{3}z_{2}-\gamma=0 
\end{array}
\right.
\]
or
\begin{equation}
\left\{
\begin{array}{ll}
\bar{\beta}z_{1}^{2}+(h_{1}-h_{3})z_{1}-\beta
=-z_{2}(\alpha +\bar{\gamma}z_{1}),  \\
\bar{\gamma}z_{2}^{2}+(h_{2}-h_{3})z_{2}-\gamma
=-z_{1}(\bar{\alpha} +\bar{\beta}z_{2}).
\end{array}
\right.
\end{equation}

\vspace{3mm}
Let us solve the equations. We get $z_{1}$ by solving the equation
\begin{eqnarray}
\label{eq:reduced-1-i}
&&\{
\bar{\beta}\bar{\gamma}(h_{1}-h_{2})
-\alpha\bar{\beta}^{2}+\bar{\alpha}\bar{\gamma}^{2}
\}z_{1}^{3}+ \nonumber \\
&&[
\bar{\gamma}
\{
(h_{1}-h_{2})(h_{1}-h_{3})+2|\alpha|^{2}-|\beta|^{2}-|\gamma|^{2}
\}
-\alpha\bar{\beta}(h_{1}+h_{2}-2h_{3})
]z_{1}^{2}+ \nonumber \\
&&[
-\alpha
\{
(h_{1}-h_{3})(h_{2}-h_{3})-|\alpha|^{2}-|\beta|^{2}+2|\gamma|^{2}
\}
+\beta\bar{\gamma}(-2h_{1}+h_{2}+h_{3})
]z_{1}+  \nonumber \\
&&\beta^{2}\bar{\gamma}+\alpha\beta(h_{2}-h_{3})-\alpha^{2}\gamma
=
0
\end{eqnarray}
by use of Cardano formula and $z_{2}$ by solving the equation
\begin{equation}
\label{eq:reduced-1-ii}
\bar{\gamma}z_{2}^{2}+(h_{2}-h_{3}+\bar{\beta}z_{1})z_{2}
+\bar{\alpha}z_{1}-\gamma=0.
\end{equation}

If $(z_{1},z_{2})$ is given then the reduced Hamiltonian becomes
\begin{eqnarray}
&&(1_{2}+Z^{\dagger}Z)^{-1/2}\widetilde{H}_{+}(1_{2}+Z^{\dagger}Z)^{-1/2}
\nonumber \\
&=&
\frac{1}{(|z_{1}|^{2}+|z_{2}|^{2})^{2}} 
\left(
\begin{array}{cc}
\frac{|z_{1}|^{2}}{\sqrt{1+|z_{1}|^{2}+|z_{2}|^{2}}}+|z_{2}|^{2}
&
\frac{\bar{z}_{1}z_{2}}{\sqrt{1+|z_{1}|^{2}+|z_{2}|^{2}}}-\bar{z}_{1}z_{2} \\
\frac{z_{1}\bar{z}_{2}}{\sqrt{1+|z_{1}|^{2}+|z_{2}|^{2}}}-z_{1}\bar{z}_{2}
&
\frac{|z_{2}|^{2}}{\sqrt{1+|z_{1}|^{2}+|z_{2}|^{2}}}+|z_{1}|^{2}
\end{array}
\right)\times  \nonumber \\
&&\quad\quad\quad\quad\quad\quad\ \
\left(
\begin{array}{cc}
h_{1}+\bar{\beta}z_{1}+\beta\bar{z}_{1}+h_{3}|z_{1}|^{2} & 
\bar{\alpha}+\gamma\bar{z}_{1}+\bar{\beta}z_{2}+h_{3}\bar{z}_{1}z_{2} \\
\alpha+\bar{\gamma}z_{1}+\beta\bar{z}_{2}+h_{3}z_{1}\bar{z}_{2} & 
h_{2}+\bar{\gamma}z_{2}+\gamma\bar{z}_{2}+h_{3}|z_{2}|^{2}
\end{array}
\right)\times  \nonumber \\
&&\quad\quad\quad\quad\quad\quad\ \
\left(
\begin{array}{cc}
\frac{|z_{1}|^{2}}{\sqrt{1+|z_{1}|^{2}+|z_{2}|^{2}}}+|z_{2}|^{2}
&
\frac{\bar{z}_{1}z_{2}}{\sqrt{1+|z_{1}|^{2}+|z_{2}|^{2}}}-\bar{z}_{1}z_{2} \\
\frac{z_{1}\bar{z}_{2}}{\sqrt{1+|z_{1}|^{2}+|z_{2}|^{2}}}-z_{1}\bar{z}_{2}
&
\frac{|z_{2}|^{2}}{\sqrt{1+|z_{1}|^{2}+|z_{2}|^{2}}}+|z_{1}|^{2}
\end{array}
\right)  \nonumber \\
&{\equiv}& 
\left(
\begin{array}{cc}
k_{1} & \bar{\zeta} \\
\zeta & k_{2}
\end{array}
\right).
\end{eqnarray}

Therefore we have only to solve the equation
\begin{equation}
\label{eq:reduced-2}
\bar{\zeta}z^{2}+(k_{1}-k_{2})z-\zeta =0
\end{equation}
as shown in the example in section 2. 

Finally we obtain three eigenvalues
\begin{eqnarray}
\lambda_{1}&=&\frac{k_{1}+\bar{\zeta}z+\zeta\bar{z}+k_{2}\zetta^{2}}
{1+\zetta^{2}},\ \ 
\lambda_{2}=\frac{k_{2}-\bar{\zeta}z-\zeta\bar{z}+k_{1}\zetta^{2}}
{1+\zetta^{2}},  \nonumber \\
\lambda_{3}&=&
\frac{h_{3}-(\bar{\beta}z_{1}+\beta\bar{z}_{1})-
(\bar{\gamma}z_{2}+\gamma\bar{z}_{2})+
h_{1}|z_{1}|^{2}+\bar{\alpha}z_{1}\bar{z}_{2}+{\alpha}\bar{z}_{1}{z}_{2}+
h_{2}|z_{2}|^{2}}
{1+|z_{1}|^{2}+|z_{2}|^{2}}
\end{eqnarray}
under (\ref{eq:reduced-1-i}),  (\ref{eq:reduced-1-ii}) and 
(\ref{eq:reduced-2}).

In the process of calculation MATHEMATICA or MAPLE is 
indispensable (calculation by force is very hard).

\section{Discussion}
In this paper we presented a geometric approach to diagonalization 
of hermitian matrices. The advantage of our method is quick 
diagonalization, while to obtain eigenvalues is left in the final step. 
Our method is in a certain sense reverse process of the standard 
one, so they are dual each other.

It is not clear at the present time whether our method is convenient 
enough or not. Further work will be needed.

Last, let us make a comment on ``Riccati structure" of Quantum 
Mechanics, see for example \cite{Mi}. 
We consider the harmonic oscillator given by
\[
H=\frac{1}{2}
\left(
-\frac{d^{2}}{dx^{2}}+x^{2}
\right).
\]
In order to solve the model we usually define the annihilation and 
creation operators like
\[
a=\frac{1}{\sqrt{2}}
\left(
\frac{d}{dx}+x
\right), \quad
a^{\dagger}=\frac{1}{\sqrt{2}}
\left(
-\frac{d}{dx}+x
\right).
\]
Then it is well--known that
\[
aa^{\dagger}=H+\frac{1}{2},\quad a^{\dagger}a=H-\frac{1}{2}.
\]

Next we define another annihilation and creation operators like
\[
b=\frac{1}{\sqrt{2}}
\left(
\frac{d}{dx}+\beta(x)
\right), \quad
b^{\dagger}=\frac{1}{\sqrt{2}}
\left(
-\frac{d}{dx}+\beta(x)
\right)
\]
with unknown $\beta(x)$.

Then, in order to satisfy the relation $bb^{\dagger}=H+\frac{1}{2}$ \ 
$\beta(x)$ must satisfy the equation
\[
\beta^{\prime}+\beta^{2}=1+x^{2}.
\]
This is a special type of Riccati differential equation.

\vspace{10mm}
\begin{center}
\begin{Large}
{\bf Appendix}
\end{Large}
\end{center}
In the appendix we consider the important example
\[
H=
\left(
\begin{array}{ccc}
h_{1} & \bar{\alpha} & \bar{\beta} \\
\alpha  & h_{2} & \bar{\gamma} \\
\beta  & \gamma & h_{3}
\end{array}
\right)
\]
once more. We want to diagonalize the matrix {\bf at a time}.

For the purpose we prepare a unitary matrix coming from 
the flag manifold of the second type (in our terminology, see  
\cite{FO}, \cite{Fu2} and also \cite{Pi}, \cite{DJ})\ 
$SU(3)/U(1)\times U(1)$ 
\begin{eqnarray}
U=U(x,y,z)&=&
\left(
\begin{array}{ccc}
1 & -(\bar{x}+\bar{y}z) & \bar{x}\bar{z}-\bar{y}  \\
x & \Delta_{1}-x(\bar{x}+\bar{y}z) & -\bar{z}      \\
y & z\Delta_{1}-y(\bar{x}+\bar{y}z) & 1
\end{array}
\right)
\left(
\begin{array}{ccc}
\frac{1}{\sqrt{\Delta_{1}}} &   &                 \\
   & \frac{1}{\sqrt{\Delta_{1}\Delta_{2}}} &   \\
   &   & \frac{1}{\sqrt{\Delta_{2}}}
\end{array}
\right) \\
&\equiv& U_{M}U_{D} \nonumber
\end{eqnarray}
where $\Delta_{1}$ and $\Delta_{2}$ are given by
\[
\Delta_{1}=1+|x|^{2}+|y|^{2},\quad
\Delta_{1}=1+|z|^{2}+|xz-y|^{2}.
\]

Let us calculate $U^{\dagger}HU$. Since\ 
$
U^{\dagger}HU=U_{D}U_{M}^{\dagger}HU_{M}U_{D}
$\ 
we have only to calculate $U_{M}^{\dagger}HU_{M}$ because 
$U_{D}$ is diagonal. Namely,
\begin{eqnarray}
U_{M}^{\dagger}HU_{M}
&=&
\left(
\begin{array}{ccc}
1 & \bar{x} & \bar{y}                                       \\
-(x+y\bar{z}) & \Delta_{1}-\bar{x}(x+y\bar{z}) & 
                 \bar{z}\Delta_{1}-\bar{y}(x+y\bar{z})  \\
xz-y & -z & 1
\end{array} 
\right)
\left(
\begin{array}{ccc}
h_{1} & \bar{\alpha} & \bar{\beta} \\
\alpha  & h_{2} & \bar{\gamma} \\
\beta  & \gamma & h_{3}
\end{array}
\right) \times  \nonumber \\
&&\left(
\begin{array}{ccc}
1 & -(\bar{x}+\bar{y}z) & \bar{x}\bar{z}-\bar{y}  \\
x & \Delta_{1}-x(\bar{x}+\bar{y}z) & -\bar{z}      \\
y & z\Delta_{1}-y(\bar{x}+\bar{y}z) & 1
\end{array}
\right)  \nonumber \\
&\equiv&
\left(
\begin{array}{ccc}
w_{11} & \bar{w}_{21} & \bar{w}_{31} \\
w_{21} & w_{22}        & \bar{w}_{32} \\
w_{31} & w_{32}        & w_{33} 
\end{array}
\right)
\end{eqnarray}
where
\begin{eqnarray}
w_{21}
&=&
-(x+y\bar{z})(h_{1}+x\bar{\alpha}+y\bar{\beta})+
\{\Delta_{1}-\bar{x}(x+y\bar{z})\}(\alpha+xh_{2}+y\bar{\gamma})+ \nonumber \\
&&\{\bar{z}\Delta_{1}-\bar{y}(x+y\bar{z})\}(\beta+x\gamma+yh_{3}), \nonumber \\
w_{31}
&=&
(xz-y)h_{1}-z\alpha+\beta+\{(xz-y)\bar{\alpha}-zh_{2}+\gamma\}x+
\{(xz-y)\bar{\beta}-z\bar{\gamma}+h_{3}\}y  \nonumber \\
&=&
(xz-y)(h_{1}+x\bar{\alpha}+y\bar{\beta})-
z(\alpha+xh_{2}+y\bar{\gamma})+(\beta+x\gamma+yh_{3}), \\
w_{32}
&=&
-(\bar{x}+\bar{y}z)[\{(xz-y)h_{1}-z\alpha+\beta\}+
\{(xz-y)\bar{\alpha}-zh_{2}+\gamma\}x+ \nonumber \\
&&\{(xz-y)\bar{\beta}-z\bar{\gamma}+h_{3}\}y]+
\Delta_{1}[(xz-y)\bar{\alpha}-zh_{2}+\gamma\}+
\{(xz-y)\bar{\beta}-z\bar{\gamma}+h_{3}\}z]  \nonumber \\
&=&
-(\bar{x}+\bar{y}z)w_{31}+
\Delta_{1}[(xz-y)\bar{\alpha}-zh_{2}+\gamma\}+
\{(xz-y)\bar{\beta}-z\bar{\gamma}+h_{3}\}z].  \nonumber
\end{eqnarray}

Here by setting $w_{21}=w_{31}=w_{32}=0$, we have
\begin{eqnarray}
0
&=&
-(x+y\bar{z})(h_{1}+x\bar{\alpha}+y\bar{\beta})+
\{\Delta_{1}-\bar{x}(x+y\bar{z})\}(\alpha+xh_{2}+y\bar{\gamma})+ \nonumber \\
&&\{\bar{z}\Delta_{1}-\bar{y}(x+y\bar{z})\}(\beta+x\gamma+yh_{3}), \nonumber \\
0
&=&
(xz-y)(h_{1}+x\bar{\alpha}+y\bar{\beta})-
z(\alpha+xh_{2}+y\bar{\gamma})+(\beta+x\gamma+yh_{3}), \\
0
&=&
(xz-y)\bar{\alpha}-zh_{2}+\gamma+
\{(xz-y)\bar{\beta}-z\bar{\gamma}+h_{3}\}z.  \nonumber
\end{eqnarray}
Some calculation gives
\begin{eqnarray}
&&
\bar{z}=-
\frac
{
x(h_{1}+x\bar{\alpha}+y\bar{\beta})-
(1+|y|^{2})(\alpha+xh_{2}+y\bar{\gamma})+
\bar{y}x(\beta+x\gamma+yh_{3})
}
{
y(h_{1}+x\bar{\alpha}+y\bar{\beta})+
\bar{x}y(\alpha+xh_{2}+y\bar{\gamma})-
(1+|x|^{2})(\beta+x\gamma+yh_{3})
},  \nonumber \\
&&
z=
\frac
{
y(h_{1}+x\bar{\alpha}+y\bar{\beta})-(\beta+x\gamma+yh_{3})
}
{
x(h_{1}+x\bar{\alpha}+y\bar{\beta})-(\alpha+xh_{2}+y\bar{\gamma})
},  \\
&&
(x\bar{\beta}-\bar{\gamma})z^{2}+(h_{3}-h_{2}+x\bar{\alpha}-
y\bar{\beta})z-y\bar{\alpha}+\gamma=0  \nonumber
\end{eqnarray}
in terms of $\Delta_{1}=1+|x|^{2}+|y|^{2}$.

It is not easy at the present time to solve the equations at a time.

%%%%%%%%%%%%%
%References%
%%%%%%%%%%%%%

\end{document}